\documentclass[preprint,aps,superscriptaddress,nofootinbib,longbibliography]{revtex4-1}
\usepackage[T1]{fontenc}
\usepackage[latin9]{inputenc}
\usepackage[active]{srcltx}
\usepackage{amstext}
\usepackage{amssymb}
\usepackage{graphicx}
\usepackage{esint}
\usepackage{babel}
\usepackage{hyperref}

\makeatletter

\providecommand{\tabularnewline}{\\}

\makeatother

\begin{document}

\title{Sphaleron portal baryogenesis}

\newcommand{\affUFABC}{Centro de Ci\^encias Naturais e Humanas\;\;\\
	Universidade Federal do ABC, 09.210-170,
	Santo Andr\'e, SP, Brazil}

\author{Chee Sheng Fong}
\email{sheng.fong@ufabc.edu.br}
\affiliation{\affUFABC}

\begin{abstract}
Nontrivial topological vacua of non-Abelian gauge symmetry $SU(3)\times SU(2)_{L}$
of the Standard Model play an important role in baryogenesis. In particular,
the baryon (and lepton) number violation from $SU(2)_{L}$ sphaleron
is a crucial ingredient for baryogenesis at weak scale or higher.
In this work, we point out that generically, a baryon asymmetry is
induced by an asymmetry generated in the new sector through strong
$SU(3)$ and/or weak $SU(2)_{L}$ sphaleron portals and vice versa.
In the standard radiation-dominated early Universe, due to phenomenological
constraints, the sphaleron portal baryogenesis has to take place at
cosmic temperature $T\gtrsim10^{6}-10^{8}$ GeV together with a $(B-L)$-violating
source. As an example, we show an explicit model where strong sphaleron
portal baryogenesis occurs at the scale of Peccei-Quinn breaking to
solve the strong CP problem and this coincides nicely with the scale
where the Weinberg operator responsible for Majorana neutrino mass
is in equilibrium.
\end{abstract}

\maketitle
\flushbottom

\section{Introduction}

Ordinary $U(1)$ global symmetries and their associated Noether charges
are useful to understand how cosmic matter-antimatter asymmetry arises
through baryogenesis \cite{Fong:2015vna}. In quantum field theory,
an exact global symmetry can arise accidentally or due to gauge symmetry.
In the Standard Model (SM), the conservation of hypercharge is due
to hypercharge gauge symmetry $U(1)_{Y}$ while the conservation of
three linear combinations of baryon $B$ and lepton flavor $L_{\alpha}$
($\alpha=e,\mu,\tau$) charges for instance, $B/3-L_{\alpha}$, arise
accidentally.\footnote{They are accidental in the sense that they can be broken by nonrenormalizable
operators.} 

In an expanding Universe, there is a new time scale associated with
the inverse of the Hubble expansion rate $t_{{\cal H}}\equiv{\cal H}^{-1}$,
and besides the aforementioned exact global symmetries, new effective
global symmetries can arise when some particle interaction rate $\Gamma_{x}$
and its associated time scale $t_{x}\equiv\Gamma_{x}^{-1}$ becomes
greater than $t_{{\cal H}}$ \cite{Fong:2010qh,Fong:2015vna}. In
Appendix \ref{app:SM}, we list the effective $U(1)_{x}$ symmetries
for the SM \cite{Domcke:2020quw,Fong:2020fwk} that arise at different
cosmic temperatures $T$ in a radiation-dominated Universe with the
Hubble rate
\begin{eqnarray}
{\cal H} & = & 1.66\sqrt{g_{\star}}\frac{T^{2}}{M_{\textrm{Pl}}},\label{eq:Hubble_rate_radiation_dominated}
\end{eqnarray}
where $M_{\textrm{Pl}}=1.22\times10^{19}$ GeV and $g_{\star}$ is
the total relativistic degrees of freedom. An effective $U(1)_{x}$
implies a conserved Noether charge $n_{x}V$ where $n_{x}$ is the
corresponding charge density and $V$ is the physical volume of the
Universe.\footnote{Assuming a homogeneous and isotropic Universe, $V$ can be any subvolume
where the assumptions hold.} These symmetries are double-edge swords in baryogenesis: on the one
hand, they prevent asymmetries from being erased while on the other
hand, they prevent asymmetries from being generated. 

For any species of particle $i$ that distinguishes from its antiparticle
$\overline{i}$, we can define its number density asymmetry as $n_{\Delta i}\equiv n_{i}-n_{\bar{i}}$.
Given a system at temperature $T$ with effective $U(1)_{x}$ symmetries
and the corresponding charge densities $n_{x}$, for particle $i$
in kinetic equilibrium, we have \cite{Fong:2015vna}
\begin{eqnarray}
n_{\Delta i} & = & g_{i}\zeta_{i}\sum_{y,x}q_{i}^{y}\left(J^{-1}\right)_{yx}n_{x},\label{eq:general_solution}
\end{eqnarray}
where we have defined
\begin{eqnarray}
\zeta_{i} & \equiv & \frac{6}{\pi^{2}}\int_{\frac{m_{i}}{T}}^{\infty}dx\,x\sqrt{x^{2}-\frac{m_{i}^{2}}{T^{2}}}\frac{e^{x}}{\left(e^{x}-\eta_{i}\right)^{2}},\label{eq:statistical_function}\\
\left(J\right)_{xy} & \equiv & \sum_{i}g_{i}\zeta_{i}q_{i}^{x}q_{i}^{y},\label{eq:J_matrix}
\end{eqnarray}
with $\eta_{i}=1\left(-1\right)$ for $i$ a boson (fermion) and $m_{i}$,
$g_{i}$ and $q_{i}^{x}$ are respectively the mass, internal degrees
of freedom and charge of particle $i$ under $U(1)_{x}$. The sum
$i$ is over all the particle types in the system. 

In the SM, both baryon $B$ and lepton $L$ numbers are violated due
to the Adler-Bell-Jackiw mixed anomaly with weak $SU(2)_{L}$ \cite{Adler:1969gk,Bell:1969ts},
resulting in the 't Hooft operator \cite{tHooft:1976rip,tHooft:1976snw}
\begin{eqnarray}
{\cal O}_{SU(2)_{L}} & = & \prod_{\alpha}\left(Q\ell\ell\ell\right)_{\alpha},\label{eq:EW_sphaleron}
\end{eqnarray}
where $Q_{\alpha}$ and $\ell_{\alpha}$ are respectively the left-handed
quark and lepton $SU(2)_{L}$ doublets with $\alpha=1,2,3$ the family
index. While the processes induced by the operator (\ref{eq:EW_sphaleron})
are exponentially suppressed at zero temperature \cite{tHooft:1976rip,tHooft:1976snw},
in a radiation-dominated Universe at temperatures $132\,\textrm{GeV}\lesssim T\lesssim10^{12}\,\textrm{GeV}$,
due to thermal effect, they proceed faster than the Hubble expansion
rate and act as the source of $B$ violation \cite{Kuzmin:1985mm}.
Although $B$ and $L$ are not part of the conserved charges of the
system, using eq. (\ref{eq:general_solution}), $n_{B}$ and $n_{L}$
can be expressed in terms of $n_{x}$ of the system. For baryogenesis,
we are interested in the baryon number which is given by
\begin{eqnarray}
n_{B} & = & \sum_{i}q_{i}^{B}n_{\Delta i}=\sum_{y,x}\left(J\right)_{By}\left(J^{-1}\right)_{yx}n_{x},\label{eq:baryon_number}
\end{eqnarray}
where in the second equality, we have used eqs. (\ref{eq:general_solution})
and the definition (\ref{eq:J_matrix}). Throughout the thermal history
of the Universe, the baryon number is proportional to the effective
charges listed in Appendix \ref{app:SM}. During this period, some
$U(1)_{x}$ have to be broken ``slowly'' i.e. in an out-of-equilibrium
fashion where the timescale of violation being $t_{x}\sim t_{{\cal H}}$
to produce a nonzero $n_{x}$, giving rise to a nonzero $n_{B}$.
In a radiation-dominated Universe with temperature $T$, its entropy
density is given by $s=\frac{2\pi^{2}}{45}g_{\star}T^{3}$. Assuming
an isentropic Universe where the entropy $sV$ is conserved\footnote{This is a very good approximation if the energy density of the Universe
is not dominated by some out-of-equilibrium fields after inflation.}, it is convenient to define ${\cal Y}_{x}\equiv n_{x}/s$ and rewrite
eq. (\ref{eq:baryon_number}) as
\begin{eqnarray}
{\cal Y}_{B} & = & \sum_{y,x}\left(J\right)_{By}\left(J^{-1}\right)_{yx}{\cal Y}_{x}.\label{eq:baryon_number_Y}
\end{eqnarray}
After genesis when all the relevant ${\cal Y}_{x}\propto n_{x}V$
become constant, ${\cal Y}_{B}$ is a constant as well. At $T\lesssim10^{4}$
GeV, the only conserved charges in the SM is the hypercharge $Y$
and $\Delta_{\alpha}\equiv B/3-L_{\alpha}\,(\alpha=1,2,3)$ and using
eq. (\ref{eq:baryon_number_Y}), we obtain
\begin{eqnarray}
{\cal Y}_{B} & = & \frac{1}{79}\left(28\sum_{\alpha}{\cal Y}_{\Delta_{\alpha}}-6{\cal Y}_{Y}\right)=\frac{1}{79}\left(28{\cal Y}_{B-L}-6{\cal Y}_{Y}\right).\label{eq:B_B-L_Y}
\end{eqnarray}
For $T<132$ GeV, the weak sphaleron processes due to operator (\ref{eq:EW_sphaleron})
will be slower than the Hubble rate, resulting in an effectively conserved
$B$ and ${\cal Y}_{B}$ will be frozen \cite{DOnofrio:2014rug}.
This value should be compared to the measured one ${\cal Y}_{B}\sim9\times10^{-11}$
\cite{Lahav:2022poa}. Considering a hypercharge neutral Universe
${\cal Y}_{Y}=0$, one concludes that $B-L$ needs to be broken to
have a nonzero ${\cal Y}_{B}$. 

From eq. (\ref{eq:baryon_number}), ref. \cite{Fong:2015vna} further
clarified that a symmetry can also act as a messenger to communicate
an asymmetry between the SM and another new sector $X$ even if it
does not carry baryon number i.e. $\left(J\right)_{BX}=0$. This intriguing
possibility was first shown in ref. \cite{Antaramian:1993nt} where
the new sector does not have to carry $B$ nor $L$ but only hypercharge.\footnote{This mechanism was also applied in ref. \cite{Servant:2013uwa} to
relate the asymmetry between a dark sector and the SM.} To see how hypercharge can act as a messenger, let us introduce a
new sector $X$ with its own conserved charge ${\cal Y}_{X}$ and
further carries a nonzero hypercharge $q_{X}^{Y}$. From eq. (\ref{eq:baryon_number_Y}),
we obtain
\begin{eqnarray}
{\cal Y}_{B} & = & \frac{1}{79}\left(28{\cal Y}_{B-L}-6{\cal Y}_{Y}+6q_{X}^{Y}{\cal Y}_{X}\right),
\end{eqnarray}
where we can have ${\cal Y}_{B-L}={\cal Y}_{Y}=0$ while ${\cal Y}_{B}\neq0$
as long as $q_{X}^{Y}{\cal Y}_{X}\neq0$. 

Ref. \cite{Antaramian:1993nt} further pointed out that even if $(B-L)$-violating
interaction is in equilibrium, baryon number can remain nonzero. Indeed,
from eq. (\ref{eq:baryon_number}), removing $\Delta_{\alpha}$ from
the list of conserved charges, we have
\begin{eqnarray}
{\cal Y}_{B} & = & \frac{2}{11}\left({\cal Y}_{Y}-q_{X}^{Y}{\cal Y}_{X}\right),
\end{eqnarray}
where with ${\cal Y}_{Y}=0$, ${\cal Y}_{B}\neq0$ as long as $q_{X}^{Y}{\cal Y}_{X}\neq0$.
This interesting fact was rediscovered in ref. \cite{AristizabalSierra:2013lyx}
pointing out that since particles in the $X$ sector must decay at
$T<132$ GeV but before the Big Bang Nucleosynthesis (BBN) $T\gtrsim$
MeV, one could have signature of hypercharged-carrying long-lived
particles in the collider. Notice that $B-L$ not being a conserved
charge, can be written in terms of ${\cal Y}_{Y}$ and ${\cal Y}_{X}$
as
\begin{eqnarray}
{\cal Y}_{B-L} & = & \frac{8}{11}\left({\cal Y}_{Y}-q_{X}^{Y}{\cal Y}_{X}\right).
\end{eqnarray}
Assuming that $(B-L)$-violating processes get out of equilibrium
at $T>132$ GeV, then ${\cal Y}_{B-L}$ will be conserved and the
final baryon number will be given by eq. (\ref{eq:B_B-L_Y}). This
type of scenario is elaborated in refs. \cite{Domcke:2020quw,Fong:2020fwk}
pointing out a new avenue for baryogenesis utilizing the many effective
symmetries that exist at high temperature in the SM itself (see Appendix
\ref{app:SM}). 

From the symmetry analysis above, it should be apparent that the existence
of effective symmetries play an important role in baryogenesis. In
particular, when considering any extension to the SM, new effective
symmetries can emerge. For instance, in order to understand why CP
is conserved in the strong sector, the most popular explanation is
to extend the SM field content such that we have an effective Peccei-Quinn
symmetry \cite{Peccei:1977hh,Peccei:1977ur} $U(1)_{PQ}$ with a $U(1)_{PQ}-SU(3)_{c}^{2}$
mixed anomaly. By construction, $U(1)_{PQ}$ will be violated by the
strong sphaleron processes at high temperature \cite{McLerran:1990de},
the strong force analog of the weak sphaleron processes described
by the operator (\ref{eq:EW_sphaleron}). Nevertheless, there exists
a linear combination of $U(1)_{PQ}$ and $U(1)_{X}$ which is effectively
conserved as first pointed out in supersymmetric context in ref. \cite{Ibanez:1992aj}
and showed to play a relevant role in supersymmetric leptogenesis
\cite{Fong:2010qh,Fong:2010bv}. In this work, we will study the implications
of the strong and weak sphaleron portals for baryogenesis.

\section{The strong sphaleron portal}

In the SM, besides the weak sphaleron operator (\ref{eq:EW_sphaleron})
which violate both $B$ and $L$ but preserve $\Delta_{\alpha}$,
we also have the strong sphaleron processes which break the quark
chiral symmetries at high temperature \cite{McLerran:1990de} and
is described by the operator
\begin{eqnarray}
{\cal O}_{SU(3)} & = & \prod_{\alpha}\left(QQU^{c}D^{c}\right)_{\alpha},\label{eq:QCD_sphaleron}
\end{eqnarray}
where $U_{\alpha}$ and $D_{\alpha}$ are respectively the up- and
down-type right-handed quarks with the subscript $c$ stands for charge
conjugation. 

When new physics degrees of freedom are added to the SM model, new
$U(1)_{X}$ symmetries can arise. As long as these new symmetries
have mixed $SU(2)_{L}$ and/or $SU(3)$ anomalies, they serve as portals
between the SM and the new physics sector. Let us define the mixed
$U(1)_{X}-SU(N)^{2}$ anomaly coefficient with $N\geq2$ as
\begin{eqnarray}
A_{XNN} & \equiv & \sum_{i}c(R_{i})g_{i}q_{i}^{X},
\end{eqnarray}
where the sum is over particle $i$ with degeneracy $g_{i}$, charge
$q_{i}^{X}$ under $U(1)_{X}$, and representation $R_{i}$ under
$SU(N)$ and the normalization is fixed as $c\left(R_{i}\right)=\frac{1}{2}$
in the fundamental representation and $c_{2}\left(R_{i}\right)=N$
in the adjoint representation. The 't Hooft operators such as those
in eqs. (\ref{eq:EW_sphaleron}) and (\ref{eq:QCD_sphaleron}) can
be constructed as follows: each fermion $i$ with representation $R_{i}$
under $SU(N)$ will contribute proportionally to $c\left(R_{i}\right)$
to the $SU(N)$ to the operator
\begin{eqnarray}
{\cal O}_{SU(N)} & \sim & \prod_{i}\Psi_{i}^{2g_{i}c_{2}\left(R_{i}\right)},\label{eq:instanton_induced_operators}
\end{eqnarray}
where the factor of 2 is included such that the field enters in integer
number. 

Let consider a $U(1)_{PQ}$ with $U(1)_{PQ}-SU(3)^{2}$ anomaly coefficient
$A_{PQ33}\neq0$. Notice that all the quark chiral symmetries are
broken by eq. (\ref{eq:QCD_sphaleron}) while the individual right-handed
quark chiral symmetries $U(1)_{q=\left\{ t,b,c,s,u,d\right\} }$ are
further broken by the quark Yukawa interactions with $U(1)_{u}$ the
last one being broken by up-quark Yukawa interactions at $T\lesssim10^{6}$
GeV. Although $U(1)_{u}$ is also broken by the operator (\ref{eq:QCD_sphaleron})
which is in equilibrium for $T\lesssim10^{13}$ GeV \cite{McLerran:1990de},
we can form a linear combination of $U(1)_{PQ}$ and $U(1)_{u}$ which
is conserved at $T\gtrsim10^{6}$ GeV. Without loss of generality,
we set the $u$ chiral charge to one such that $A_{u33}=\frac{1}{2}$.
Then the new conserved charge at $T\gtrsim10^{6}$ GeV is
\begin{eqnarray}
\overline{PQ} & \equiv & -\frac{1}{2}\frac{PQ}{A_{PQ33}}+u.\label{eq:PQ_bar}
\end{eqnarray}
From eq. (\ref{eq:baryon_number_Y}), we will have ${\cal Y}_{B}\propto{\cal Y}_{\overline{PQ}}$.
Hence strong sphaleron portal baryongesis can proceed from the slow
breaking of $\overline{PQ}$ which can come from the spontaneous breaking
of $\overline{PQ}$ itself as we will show in Section \ref{sec:model}.
Another interesting possibility proposed in ref. \cite{Co:2019wyp}
is to induce a $PQ$ charge through the rotation of axion field due
to higher dimensional $PQ$-breaking operators. For temperature below
\begin{eqnarray}
T_{\overline{PQ}} & \sim & 10^{6}\,\textrm{GeV},\label{eq:T_PQbar}
\end{eqnarray}
due to the up quark Yukawa interactions, $\overline{PQ}$ is no longer
conserved and ${\cal Y}_{B}$ will be driven to zero. This conclusion
can be circumvented in nonstandard cosmology if the cosmic expansion
rate is much faster as predicted in a class of scalar-tensor theories
of gravitation \cite{Dutta:2018zkg} and will be explored elsewhere. 

Another elegant option is to induce nonzero $B-L$ charge during the
$\overline{PQ}$-genesis by considering a $(B-L)$-violating source
which is in thermal equilibrium \cite{Domcke:2020quw,Fong:2020fwk}.
One can connect this to Majorana neutrino mass which arises from the
$(B-L)$-violating Weinberg operator \cite{Weinberg:1979sa,Weinberg:1980bf}
\begin{eqnarray}
{\cal O}_{5} & \equiv & \frac{c_{\alpha\beta}}{\Lambda}\ell_{\alpha}H\ell_{\beta}H,\label{eq:Weinberg_operator}
\end{eqnarray}
where the neutrino mass matrix is $m_{\nu}=cv^{2}/\Lambda$ with the
SM Higgs $H$ vacuum expectation value (vev) $v=\left\langle H\right\rangle =174$
GeV. For $T<\Lambda$, by comparing the Hubble rate in eq. (\ref{eq:Hubble_rate_radiation_dominated}),
the $(B-L)$-violating interaction mediated by the Weinberg operator
$\Gamma_{B-L}\sim\left|m_{\nu}\right|^{2}T^{3}/v^{4}$ is in equilibrium
if $T$ is greater than
\begin{eqnarray}
T_{B-L} & \sim & 10^{11}\,{\rm GeV}\left(\frac{0.1\,{\rm eV}}{\left|m_{\nu}\right|}\right)^{2},\label{eq:T_B-L}
\end{eqnarray}
where $\left|m_{\nu}\right|$ indicates the largest element in the
mass matrix. In this case, we will have ${\cal Y}_{B-L}\propto{\cal Y}_{\overline{PQ}}$
which will be frozen for $T\lesssim T_{B-L}$. While baryogenesis
through leptogenesis \cite{Fukugita:1986hr,Fong:2012buy} can occur
through some UV completions of (\ref{eq:Weinberg_operator}), if the
Universe has a maximum temperature $T<\Lambda$ after inflation, leptogenesis
is not efficient.

Next, let us consider additional $PQ$-breaking interactions that
can erase the asymmetry. There are only two options to have an $U(1)_{X}-SU(3)^{2}$
mixed anomaly (see Appendix \ref{app:SM}): i) by coupling up- and
down-type quarks to two different Higgs doublets such as in the original
Peccei-Quinn-Weinberg-Wilczek (PQWW) model \cite{Peccei:1977hh,Peccei:1977ur,Weinberg:1977ma,Wilczek:1977pj}
or its extension Dine-Fischler-Srednicki-Zhitnitsk (DFSZ) model \cite{Zhitnitsky:1980tq,Dine:1981rt}
or ii) by introducing new colored fermion(s) as in the Kim-Shifman-Vainshtein-Zakharov
(KSVZ) model \cite{Kim:1979if,Shifman:1979if}. 

For the first option, let us introduce two Higgs doublets $H_{u}$
and $H_{d}$ with opposite hypercharges which couple to the up-type
and down-type Yukawa interactions, respectively (for further discussion, see Appendix \ref{app:2HDM}). Let us consider the
following term which explicitly breaks $PQ$ and $\overline{PQ}$
\begin{eqnarray}
{\cal O}_{PQ1} & \equiv & \mu^{2}H_{u}H_{d}.
\end{eqnarray}
This term can be generated through spontaneous breaking of $U(1)_{PQ}$.
Estimating the interaction rate as $\Gamma_{PQ1}\sim\mu^{4}/T^{3}$
and comparing with the Hubble rate in eq. (\ref{eq:Hubble_rate_radiation_dominated}),
the process will be in equilibrium for temperature smaller than
\begin{eqnarray}
T_{PQ1} & \sim & 10^{5}\,\textrm{GeV}\left(\frac{\mu}{100\,\textrm{GeV}}\right)^{4/5}\left(\frac{110.75}{g_{\star}}\right)^{1/10},
\end{eqnarray}
where a phenomenological viable reference value of $\mu$ is chosen
since it contributes to the charged Higgs masses.

For the second option, the minimal option will be to introduce a pair
of colored fermions ${\cal Q}_{L}\sim\left(3,1,y_{{\cal Q}}\right)$
and ${\cal Q}_{R}\sim\left(3,1,y_{{\cal Q}}\right)$ under the SM
gauge group $SU(3)\times SU(2)_{L}\times U(1)_{Y}$ (for further discussion, see Appendix \ref{app:SU3_anomalous_charge}). In this case,
we should consider
\begin{eqnarray}
{\cal O}_{PQ2} & \equiv & m_{{\cal Q}}\overline{{\cal Q}_{R}}{\cal Q}_{L}.
\end{eqnarray}
Again this term can arise through spontaneous breaking of $U(1)_{PQ}$.
Estimating the interaction rate as $\Gamma_{PQ2}\sim m_{{\cal Q}}^{2}/T$
and comparing with the Hubble rate as before, the processes are in
equilibrium for temperature below
\begin{eqnarray}
T_{PQ2} & \sim & 9\times10^{7}\,\textrm{GeV}\left(\frac{m_{{\cal Q}}}{1\,\textrm{TeV}}\right)^{2/3}\left(\frac{106.75}{g_{\star}}\right)^{1/6},
\end{eqnarray}
where a phenomenological viable reference value of $m_{{\cal Q}}$
is chosen for the nonobservation of new colored states at particle
colliders. Altogether, in the standard radiation-dominated Universe,
strong sphaleron portal baryogenesis should occur at $T\gtrsim10^{6}-10^{8}$
GeV.

\section{The weak sphaleron portal}

Similar to the strong sphaleron portal, we will proceed by considering
a $U(1)_{X}$ with $U(1)_{X}-SU(2)_{L}^{2}$ anomaly coefficient $A_{X22}\neq0$.
Since $B$ and $L$ have $A_{B22}=A_{L22}=\frac{3}{2}$, we can always
construct an anomaly-free charge \cite{Ibanez:1992aj,Fong:2015vna}
\begin{eqnarray}
\overline{X} & \equiv & -\frac{3}{2}\frac{X}{A_{X22}}+\frac{c_{B}B+c_{L}L}{c_{B}+c_{L}},\label{eq:X_bar}
\end{eqnarray}
which is conserved at all temperature. 

In order to have an $U(1)_{X}-SU(2)_{L}^{2}$ anomaly beyond those
of $B$ and $L$, we have to introduce new fermion(s) with nontrivial
representation under $SU(2)_{L}$ (see Appendix \ref{app:SM} and \ref{app:SU2_anomalous_charge}). The
minimal option will be to introduce a pair of fermions $\psi_{L}\sim\left(1,2,y_{\psi}\right)$
and $\psi_{R}\sim\left(1,2,y_{\psi}\right)$ with the mass term
\begin{eqnarray}
{\cal O}_{\psi} & \equiv & m_{{\cal \psi}}\overline{\psi_{R}}{\cal \psi}_{L}.
\end{eqnarray}
Depending on the hypercharge $y_{\psi}$, a conservative constraint
will be $m_{\psi}\gtrsim100$ GeV. Estimating the $\overline{X}$-violating
interaction rate as $\Gamma_{\psi}\sim m_{\psi}^{2}/T$ and comparing
with the Hubble rate in eq. (\ref{eq:Hubble_rate_radiation_dominated}),
the processes are in equilibrium for temperature below
\begin{eqnarray}
T_{\overline{X}} & \sim & 2\times10^{7}\,\textrm{GeV}\left(\frac{m_{\psi}}{100\,\textrm{GeV}}\right)^{2/3}\left(\frac{106.75}{g_{\star}}\right)^{1/6},
\end{eqnarray}
So we conclude in the standard radiation-dominated Universe, weak
sphaleron portal baryogenesis should occur at $T\gtrsim10^{7}$ GeV.

\section{A model\label{sec:model}}

\begin{figure}
\begin{centering}
\includegraphics[scale=0.6]{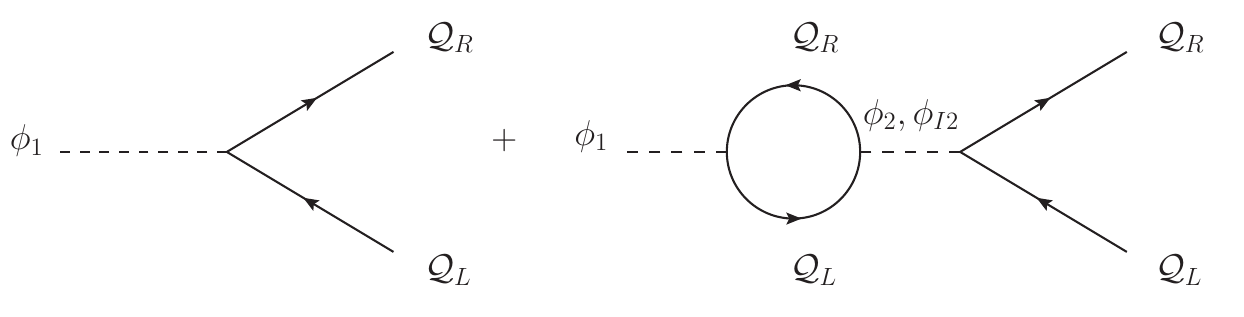}
\par\end{centering}
\caption{CP violation from the decays $\phi_{1}\to{\cal Q}_{R}+\overline{{\cal Q}_{L}}$
and its conjugate process $\phi_{1}\to\overline{{\cal Q}_{R}}+{\cal Q}_{L}$
(not shown).\label{fig:CP-violation}}

\end{figure}

Let us consider a KSVZ-type model where baryogenesis can occur through
the strong sphaleron portal. In this model, the SM is extended with
a colored fermion pair ${\cal Q}_{L}\sim\left(3,1,y_{{\cal Q}}\right)$,
${\cal Q}_{R}\sim\left(3,1,y_{{\cal Q}}\right)$ and two scalar fields
$\Phi_{i}\sim\left(1,1,0\right)$ ($i=1,2$) with the following Yukawa
interactions
\begin{eqnarray}
-{\cal L} & \supset & y_{{\cal Q}i}\overline{{\cal Q}_{R}}{\cal Q}_{L}\Phi_{i}+\textrm{H.c.},\label{eq:model}
\end{eqnarray}
and in the scalar potential, we allow all terms involving the combinations $\Phi_i \Phi_j^\dagger$ but forbid those with $\Phi_i \Phi_j$ for all $i,j=1,2$ such that $q_{\Phi_1}^X = q_{\Phi_2}^X$.\footnote{This can be achieved for example with a discrete symmetry \cite{Lazarides:1982tw}.}
There is an exact $U(1)_{{\cal Q}}$ with $q_{{\cal Q}_{L}}^{{\cal Q}}=q_{{\cal Q}_{R}}^{{\cal Q}}$
and another anomalous $U(1)_{PQ}$ with $q_{{\cal Q}_{L}}^{{\cal PQ}}-q_{{\cal Q}_{R}}^{{\cal PQ}}=-q_{\Phi_{i}}^{{\cal PQ}}$
and $A_{PQ33}=\frac{1}{2}\left(q_{{\cal Q}_{L}}^{PQ}-q_{{\cal Q}_{R}}^{PQ}\right)=-\frac{1}{2}q_{\Phi_{i}}^{PQ}\neq0$.
In particular, the strong sphaleron operator (\ref{eq:QCD_sphaleron})
is modified to
\begin{eqnarray}
{\cal O}_{SU(3),{\cal Q}} & = & {\cal Q}_{L}{\cal Q}_{R}^{c}\prod_{\alpha}\left(QQU^{c}D^{c}\right)_{\alpha}.\label{eq:SU3_Q_sphaleron}
\end{eqnarray}
 From eq. (\ref{eq:PQ_bar}), the anomaly-free charge is
\begin{eqnarray}
\overline{PQ} & = & \frac{PQ}{q_{\Phi_{i}}^{PQ}}+u.
\end{eqnarray}
Next, let us consider the spontaneous breaking of $U(1)_{\overline{PQ}}$
when $\Phi_{i}$ acquire vevs. Without loss of generality, by choosing
an appropriate linear combination of the scalar fields\footnote{For example, considering the original basis where $\widetilde \Phi_1 = v_1 + \frac{1}{\sqrt{2}}(\widetilde \phi_1 + i \widetilde \phi_{I1})$ and $\widetilde \Phi_1 = v_2 + \frac{1}{\sqrt{2}}(\widetilde \phi_2 + i \widetilde \phi_{I2})$, we can construct $\Phi_1 = \frac{1}{v_\Phi}(v_1\widetilde \Phi_1 + v_2 \widetilde \Phi_1)$ and $\Phi_2 = \frac{1}{v_\Phi}(v_2\widetilde \Phi_1 - v_1 \widetilde \Phi_1)$ where only $\Phi_1$ gets the vev $v_\Phi = \sqrt{v_1^2 + v_2^2}$.}, we can take $\left\langle \Phi_{1}\right\rangle \equiv v_{\Phi}$, $\left\langle \Phi_{2}\right\rangle = 0$
and then ${\cal Q}_{L,R}$ acquires a Dirac mass $m_{{\cal Q}}=y_{{\cal Q}}v_{\Phi}$.
Considering $\Phi_{1}=v_{\Phi}+\frac{1}{\sqrt{2}}\left(\phi_{1}+i\phi_{I1}\right)$
and $\Phi_{2}=\frac{1}{\sqrt{2}}\left(\phi_{2}+i\phi_{I2}\right)$
we have
\begin{eqnarray}
-{\cal L} & \supset & m_{{\cal Q}}\overline{{\cal Q}_{R}}{\cal Q}_{L}+\frac{1}{\sqrt{2}}y_{{\cal Q}i}\overline{{\cal Q}_{R}}{\cal Q}_{L}\left(\phi_{i}+i\phi_{Ii}\right)+\textrm{H.c.}.
\end{eqnarray}
Here $\phi_{I1}$ is the massless Nambu Goldstone boson while $\phi_{1}$,
$\phi_{2}$ and $\phi_{I2}$ have masses $m_{1}$, $m_{2}$ and $m_{I2}$.
Considering $m_{{\cal Q}}\ll m_{1}<m_{2}=m_{2I}$, a $\overline{PQ}$
symmetry can be generated from the decays $\phi_{1}\to{\cal Q}_{R}+\overline{{\cal Q}_{L}}$
and $\phi_{1}\to\overline{{\cal Q}_{R}}+{\cal Q}_{L}$ where the CP
violation arises at one-loop level as shown in Figure \ref{fig:CP-violation}.
We can quantify the CP violation by
\begin{eqnarray}
\epsilon & \equiv & \frac{\Gamma\left(\phi_{1}\to{\cal Q}_{R}+\overline{{\cal Q}_{L}}\right)-\Gamma\left(\phi_{1}\to\overline{{\cal Q}_{R}}+{\cal Q}_{L}\right)}{\Gamma\left(\phi_{1}\to{\cal Q}_{R}+\overline{{\cal Q}_{L}}\right)+\Gamma\left(\phi_{1}\to\overline{{\cal Q}_{R}}+{\cal Q}_{L}\right)}\simeq\frac{3\textrm{Im}\left[\left(y_{{\cal Q}1}y_{{\cal Q}2}^{*}\right)^{2}\right]}{16\pi\left|y_{{\cal Q}1}\right|^{2}}\frac{1}{m_{2}^{2}/m_{1}^{2}-1},
\end{eqnarray}
where we have ignored $m_{{\cal Q}}$. The $\overline{PQ}$ asymmetry
generated can be expressed as
\begin{eqnarray}
{\cal Y}_{\overline{PQ}} & = & \epsilon\,\eta{\cal Y}_{\phi_{1}}^{\textrm{eq},0},
\end{eqnarray}
where ${\cal Y}_{\phi_{1}}^{\textrm{eq},0}\equiv\frac{45\zeta\left(3\right)}{2\pi^{4}g_{\star}}=0.278/g_{\star}$
is the relativistic $\phi_{1}$ abundance and $\eta$ is a dimensionless
number taking into account the initial abundance of $\phi_{1}$ as
well as the washout effects.

In the absence of $(B-L)$-violating interaction, using the effective
symmetries at $T\sim10^{11}$ GeV \cite{Fong:2020fwk} (see Appendix
\ref{app:SM}), we have
\begin{eqnarray}
{\cal Y}_{B} & = & \frac{576}{949}{\cal Y}_{\overline{PQ}}={\cal Y}_{L},
\end{eqnarray}
where the result is independent of the choice of $q_{{\cal Q}_{L}}^{PQ}$,
$q_{{\cal Q}_{R}}^{PQ}$ and $y_{{\cal Q}}$. Since the choice of
global charge is a convention, physics must be independent of $q_{{\cal Q}_{L}}^{PQ}$,
$q_{{\cal Q}_{R}}^{PQ}$ subject to only $q_{{\cal Q}_{L}}^{{\cal PQ}}-q_{{\cal Q}_{R}}^{{\cal PQ}}=-q_{\Phi_{i}}^{{\cal PQ}}$.
The independence of gauge charge $y_{{\cal Q}}$ is more subtle. It
is due to the vectorlike nature where both ${\cal Q}_{L}$ and ${\cal Q}_{R}$
carry the same $y_{{\cal Q}}$ and the conservation of $U(1)_{{\cal Q}}$.
In fact, the conservation of $U(1)_{{\cal Q}}$ depends on the choice
$y_{{\cal Q}}$. In the original KSVZ model where $y_{{\cal Q}}=0$,
${\cal Q}_{L,R}$ are absolutely stable since they cannot couple to
any of the SM fields and could constitute colored dark matter \cite{DeLuca:2018mzn}.
To alleviate cosmological constraints of stable colored matter, refs.
\cite{DiLuzio:2016sbl,DiLuzio:2017pfr} considered various $y_{{\cal Q}}\neq0$
in which ${\cal Q}_{L,R}$ can decay before the BBN.

As discussed earlier around eq. (\ref{eq:T_PQbar}), ${\cal Y}_{B}$
will not survive below $T_{\overline{PQ}}\sim10^{6}$ GeV in the standard
radiation-dominated early Universe. Considering the Weinberg operator
in eq. (\ref{eq:Weinberg_operator}) as the $(B-L)$-violating source
and requiring $\phi_{1}$ to decay at $T\sim m_{1}\gtrsim T_{B-L}$,
we have
\begin{eqnarray}
\left|y_{{\cal Q}1}\right| & \gtrsim & 10^{-3}\left(\frac{g_{\star}}{112}\right)^{1/4}\left(\frac{T_{B-L}}{10^{11}\,{\rm GeV}}\right)^{1/2}\sim10^{-3}\left(\frac{g_{\star}}{112}\right)^{1/4}\left(\frac{0.1\,{\rm eV}}{\left|m_{\nu}\right|}\right),
\end{eqnarray}
where in the latter relation we have used eq. (\ref{eq:T_B-L}). Removing
$B-L$ from the list of conserved charges at $T\sim10^{11}$ GeV,
we obtain
\begin{eqnarray}
{\cal Y}_{B-L} & = & \frac{6}{13}{\cal Y}_{\overline{PQ}},
\end{eqnarray}
and the final baryon number is given by eq. (\ref{eq:B_B-L_Y}). For
instance, taking $m_{2}^{2}/m_{1}^{2}\sim9$, the final baryon asymmetry
is
\begin{eqnarray}
{\cal Y}_{B} & \sim & 9\times10^{-11}\left(\frac{\eta}{0.3}\right)\left(\frac{\left|y_{{\cal Q}2}\right|\sin2\phi_{12}}{10^{-2}}\right)^{2}\left(\frac{112}{g_{\star}}\right),
\end{eqnarray}
where we have defined the CP phase by $y_{{\cal Q}1}y_{{\cal Q}2}^{*}\equiv\left|y_{{\cal Q}1}y_{{\cal Q}2}^{*}\right|e^{i\phi_{12}}$.
Although we introduce one additional scalar field than the original
KSVZ model, the solution to the strong CP problem remains unchanged.
If one is not concerned about solving the strong CP problem, explicit
$PQ$-violating source can be added for the purpose of baryogenesis.
Similar construction can be carried out for weak sphaleron portal
baryogenesis using the new charge $\overline{X}$.

\section{Discussions}

Besides the well-known role of weak sphaleron interactions as the
source of baryon number violation for baryogenesis, we have shown
that both strong and weak sphaleron interactions can act as portals
for baryogenesis. Through these portals, an asymmetry induced in the
new sector will necessarily induce a nonzero baryon number. We have
identified the strong and weak portal charges $\overline{PQ}$ and
$\overline{X}$ defined respectively in (\ref{eq:PQ_bar}) and (\ref{eq:X_bar}).
In the standard radiation-dominated early Universe, $\overline{PQ}$
will be broken by up-quark Yukawa interactions at $T\lesssim10^{6}$
GeV while $\overline{X}$ remains intact at all temperatures. Nevertheless,
once we take into account phenomenological constraints on the new
degrees of freedom required for sphaleron portal baryogenesis, we
conclude that the mechanism should occur at $T\gtrsim10^{6}-10^{8}$
GeV together with some $(B-L)$-violating source. A deviation from
the standard cosmology can allow to remove these two requirements
and will be considered elsewhere. In the KSVZ-type model that we have
considered, the scale of strong portal baryogenesis is the scale where
$PQ$ is spontaneously broken. There is a remarkable coincidence that
the required scale to solve the strong CP is just where the Weinberg
operator for neutrino mass is in equilibrium.

As a final remark, while we have focused on asymmetry generated in
a new sector which then induce a nonzero baryon number, the sphaleron
portals clearly act both ways. Baryogenesis in the SM sector will
necessarily induce an asymmetry in the new sector through the sphaleron
portals. While the new sector has to involve new degrees of freedom
that carry either color charge and/or weak charge, the lightest electrically
charged neutral can constitute (asymmetric) dark matter and one might
be able to establish a deeper connection between the visible and dark
sectors.

\begin{acknowledgments}
	C.S.F. thanks Andre Lessa for discussions and Enrico Nardi for comments on the manuscript. This work is supported by grant 2019/11197-6 and 2022/00404-3 from São Paulo Research Foundation (FAPESP).
\end{acknowledgments}

\appendix

\section{The Standard Model\label{app:SM}}

In the SM, the Yukawa interactions are described by
\begin{eqnarray}
-{\cal L}_{\textrm{SM}} & \supset & y_{u}\overline{U}QH+y_{d}\overline{D}QH^{*}+y_{e}\overline{E}\ell H^{*}+\textrm{H.c.},\label{eq:SM_Yukawas}
\end{eqnarray}
where $Q$ and $\ell$ are respectively the left-handed quark and
lepton $SU(2)_{L}$ doublets, $U$, $D$ and $E$ are respectively
the right-handed up-type quark, down-type quark and lepton $SU(2)_{L}$
singlets, and $H$ is the SM Higgs. Altogether, there are six types
of fields and requiring that the three Yukawa interactions above conserve
an arbitrary $X$ charge\footnote{Here we have suppressed the family indices of quarks and leptons assuming
that they carry family-independent charges.}
\begin{eqnarray}
q_{Q}^{X}-q_{U}^{X}+q_{H}^{X} & = & 0,\quad q_{Q}^{X}-q_{D}^{X}-q_{H}^{X}=0,\quad q_{\ell}^{X}-q_{E}^{X}-q_{H}^{X}=0,\label{eq:SM_Yukawa_conditions}
\end{eqnarray}
we obtain three classical charges: hypercharge $Y$, baryon number
$B$ and lepton number $L$ (or their linear combinations). As is
well-known, only $Y$ is anomaly-free while $B$ and $L$ have mixed
$SU(2)_{L}$ anomalies. On top of the charge conservation conditions
from the Yukawa interactions above, if we impose two additional mixed
$U(1)_{X}-SU(2)_{L}^{2}$ and $U(1)_{X}-SU(3)^{2}$ anomaly-free conditions
\begin{eqnarray}
U(1)_{X}-SU(2)_{L}^{2} & : & 3q_{Q}^{X}+q_{\ell}^{X}=0,\label{eq:SU2_anomaly_condition}\\
U(1)_{X}-SU(3)^{2} & : & 2q_{Q}^{X}-q_{U}^{X}-q_{D}^{X}=0,\label{eq:SU3_anomaly_condition}
\end{eqnarray}
the solution is
\begin{eqnarray}
q_{H}^{X} & = & -\frac{1}{4}\left(q_{E}^{X}+3q_{D}^{X}\right),\; q_{\ell}^{X}=-3q_{Q}^{X}=\frac{3}{4}\left(q_{E}^{X}-q_{D}^{X}\right),\; q_{U}^{X}=-\frac{1}{2}\left(q_{E}^{X}+q_{D}^{X}\right).
\end{eqnarray}
It is interesting to note while $U(1)_{X}-SU(2)_{L}^{2}$ anomaly-free
condition is an independent condition, $U(1)_{X}-SU(3)^{2}$ anomaly-free
condition is \emph{redundant} since it is implied by the up- and down-type
quark Yukawa interactions in eq. (\ref{eq:SM_Yukawa_conditions}).
With six fields and four conditions, we have only two independent
charges: $Y$ and $B-L$ which are given by conditions $q_{E}^{X}-3q_{D}^{X}=0$
and $q_{E}^{X}+3q_{D}^{X}=0$, respectively.

From the analysis above, we conclude that:
\begin{enumerate}
\item To have an $U(1)_{X}-SU(3)^{2}$ anomaly, we have two options: i)
by coupling up- and down-type quarks to two different Higgs doublets
as we will discuss in Appendix \ref{eq:2HD}; ii) by introducing new
colored fermion(s).
\item To have an $U(1)_{X}-SU(2)_{L}^{2}$ anomaly beyond that of $B$ and
$L$, we have one option: by introducing new fermion(s) with nontrivial
representation under $SU(2)_{L}$.
\end{enumerate}
Denoting $q_{E}^{X}-3q_{D}^{X}=a$ and $q_{E}^{X}+3q_{D}^{X}=b$,
we can write the solution as
\begin{eqnarray}
q_{H}^{X} & = & -\frac{b}{4},\; q_{\ell}^{X}=-3 q_{Q}^{X}=\frac{2a+b}{4},\; q_{E}^{X}=\frac{a+b}{2},\; q_{U}^{X}=-\frac{a+2b}{6},\; q_{D}^{X}=-\frac{a-b}{6}.\label{eq:anomaly-free_solution}
\end{eqnarray}
Any solution with $a\neq0$ and $b\neq0$ correspond to some linear
combination of $Y$ and $B-L$. With $a=0$ and $b\neq0$, we have
\begin{eqnarray}
q_{H}^{Y} & = & -\frac{b}{4},\quad q_{\ell}^{Y}=-3q_{Q}^{Y}=\frac{b}{4},\quad q_{E}^{Y}=\frac{b}{2},\quad q_{U}^{Y}=-\frac{b}{3},\quad q_{D}^{Y}=\frac{b}{6},
\end{eqnarray}
while with $b=0$ and $a\neq0$, we have
\begin{eqnarray}
q_{H}^{B-L} & = & 0,\quad q_{\ell}^{B-L}=-3q_{Q}^{B-L}=\frac{a}{2},\quad q_{E}^{B-L}=\frac{a}{2},\quad q_{U}^{B-L}=-\frac{a}{6},\quad q_{D}^{B-L}=-\frac{a}{6}.
\end{eqnarray}
Explicitly, a linear combination with no mixed anomaly can be written
as
\begin{eqnarray}
X & = & c_{Y}Y+c_{B-L}B-L,
\end{eqnarray}
where $c_{Y}$ and $c_{B-L}$ are any numbers and the $X$ charges
of the fields can be obtained from eq. (\ref{eq:anomaly-free_solution})
with the replacements $b\to c_{Y}b$ and $a\to c_{B-L}a$.

In the supersymmetrized version of the SM, we can add three more Yukawa
interactions $U^{c}D^{c}D^{c}$, $QD^{c}\ell$ and $\ell\ell E^{c}$
in the superpotential which break $B-L$. Holomorphicity requires
us to add one more left-handed Higgs chiral superfields. By identifying
$H^{*}\to H_{d}$ and $H\to H_{u}$ in eq. (\ref{eq:SM_Yukawas}),
we can further introduce a $\mu H_{u}H_{d}$ term. Notice that $\ell\ell E^{c}$,
$QD^{c}\ell$ and $E^{c}\ell H_{d}$ imply $D^{c}QH_{d}$ and we have
one less condition. Altogether, with seven fields and six conditions,
the only solution is the anomaly-free $Y$.

Once we take into account the expansion of the Universe, the situation
becomes much more interesting. One can gain (lose) an effective $U(1)$
symmetry as the interactions become slower (faster) than the Hubble
expansion rate \cite{Fong:2010qh,Fong:2015vna}. In this case, quarks
and leptons of different families are differentiated due to differences
in the Yukawa couplings. Assuming a radiation-dominated primordial
Universe, we have the effective $U(1)_{x}$ symmetries listed in Table
\ref{tab:SM_eff_symmetries} which will be broken at $T_{x}$ as we
go down in cosmic temperature \cite{Fong:2020fwk}
\begin{eqnarray}
T_{t} & \sim & 10^{15}\,{\rm GeV},\nonumber \\
T_{u} & \sim & 2\times10^{13}\,{\rm GeV},\nonumber \\
T_{B} & \sim & 2\times10^{12}\,{\rm GeV},\nonumber \\
T_{\tau} & \sim & 4\times10^{11}\,{\rm GeV},\nonumber \\
T_{u-b} & \sim & 3\times10^{11}\,{\rm GeV},\nonumber \\
T_{u-c} & \sim & 2\times10^{10}\,{\rm GeV},\\
T_{\mu} & \sim & 10^{9}\,{\rm GeV},\nonumber \\
T_{B_{3}-B_{2}} & \sim & 9\times10^{8}\,{\rm GeV}\nonumber \\
T_{u-s} & \sim & 3\times10^{8}\,{\rm GeV},\nonumber \\
T_{B_{3}+B_{2}-2B_{1}} & \sim & 10^{7}\,{\rm GeV},\nonumber \\
T_{u-d} & \sim & 2\times10^{6}\,{\rm GeV},\nonumber \\
T_{e} & \sim & 3\times10^{4}\,{\rm GeV}.\nonumber 
\end{eqnarray}
While $U(1)_{u}$ and $U(1)_{B}$ are broken respectively by the $SU(3)$
and $SU(2)$ sphaleron interactions, the rest are broken by the Yukawa
interactions. Eventually, the hypercharge gauge symmetry $Y$ is broken
at $T_{\textrm{EW}}\sim132$ GeV from the electroweak symmetry breaking
\cite{DOnofrio:2014rug} while there remain three exactly conserved
charges $\Delta_{\alpha}\equiv\frac{B}{3}-L_{\alpha}$ with $L_{\alpha}$
the lepton flavor number. 

\begin{table}
\begin{centering}
\begin{tabular}{|c|c|c|c|c|c|c|c|c|c|c|c|c|c|c|c|c|}
\hline 
$x$ & $Q_{3}$ & $Q_{2}$ & $Q_{1}$ & $U_{3}$ & $U_{2}$ & $U_{1}$ & $D_{3}$ & $D_{2}$ & $D_{1}$ & $\ell_{3}$ & $\ell_{2}$ & $\ell_{3}$ & $E_{3}$ & $E_{2}$ & $E_{1}$ & $H$\tabularnewline
\hline 
\hline 
$t$ & $0$ & $0$ & $0$ & $1$ & $0$ & $0$ & $0$ & $0$ & $0$ & $0$ & $0$ & $0$ & $0$ & $0$ & $0$ & $0$\tabularnewline
\hline 
$u$ & $0$ & $0$ & $0$ & $0$ & $0$ & $1$ & $0$ & $0$ & $0$ & $0$ & $0$ & $0$ & $0$ & $0$ & $0$ & $0$\tabularnewline
\hline 
$B$ & $\frac{1}{3}$ & $\frac{1}{3}$ & $\frac{1}{3}$ & $\frac{1}{3}$ & $\frac{1}{3}$ & $\frac{1}{3}$ & $\frac{1}{3}$ & $\frac{1}{3}$ & $\frac{1}{3}$ & $0$ & $0$ & $0$ & $0$ & $0$ & $0$ & $0$\tabularnewline
\hline 
$\tau$ & $0$ & $0$ & $0$ & $0$ & $0$ & $0$ & $0$ & $0$ & $0$ & $0$ & $0$ & $0$ & $1$ & $0$ & $0$ & $0$\tabularnewline
\hline 
$u-b$ & $0$ & $0$ & $0$ & $0$ & $0$ & $1$ & $-1$ & $0$ & $0$ & $0$ & $0$ & $0$ & $0$ & $0$ & $0$ & $0$\tabularnewline
\hline 
$u-c$ & $0$ & $0$ & $0$ & $0$ & $-1$ & $1$ & $0$ & $0$ & $0$ & $0$ & $0$ & $0$ & $0$ & $0$ & $0$ & $0$\tabularnewline
\hline 
$\mu$ & $0$ & $0$ & $0$ & $0$ & $0$ & $0$ & $0$ & $0$ & $0$ & $0$ & $0$ & $0$ & $0$ & $1$ & $0$ & $0$\tabularnewline
\hline 
$B_{3}-B_{2}$ & $\frac{1}{3}$ & $-\frac{1}{3}$ & $0$ & $\frac{1}{3}$ & $-\frac{1}{3}$ & $0$ & $\frac{1}{3}$ & $-\frac{1}{3}$ & $0$ & $0$ & $0$ & $0$ & $0$ & $0$ & $0$ & $0$\tabularnewline
\hline 
$u-s$ & $0$ & $0$ & $0$ & $0$ & $0$ & $1$ & $0$ & $-1$ & $0$ & $0$ & $0$ & $0$ & $0$ & $0$ & $0$ & $0$\tabularnewline
\hline 
$B_{3}+B_{2}-2B_{1}$ & $\frac{1}{3}$ & $\frac{1}{3}$ & $-\frac{2}{3}$ & $\frac{1}{3}$ & $\frac{1}{3}$ & $-\frac{2}{3}$ & $\frac{1}{3}$ & $\frac{1}{3}$ & $-\frac{2}{3}$ & $0$ & $0$ & $0$ & $0$ & $0$ & $0$ & $0$\tabularnewline
\hline 
$u-d$ & $0$ & $0$ & $0$ & $0$ & $0$ & $1$ & $0$ & $0$ & $-1$ & $0$ & $0$ & $0$ & $0$ & $0$ & $0$ & $0$\tabularnewline
\hline 
$e$ & $0$ & $0$ & $0$ & $0$ & $0$ & $0$ & $0$ & $0$ & $0$ & $0$ & $0$ & $0$ & $0$ & $0$ & $1$ & $0$\tabularnewline
\hline 
$Y$ & $\frac{1}{6}$ & $\frac{1}{6}$ & $\frac{1}{6}$ & $\frac{2}{3}$ & $\frac{2}{3}$ & $\frac{2}{3}$ & $-\frac{1}{3}$ & $-\frac{1}{3}$ & $-\frac{1}{3}$ & $-\frac{1}{2}$ & $-\frac{1}{2}$ & $-\frac{1}{2}$ & $-1$ & $-1$ & $-1$ & $\frac{1}{2}$\tabularnewline
\hline 
$\Delta_{3}$ & $\frac{1}{9}$ & $0$ & $0$ & $\frac{1}{9}$ & $0$ & $0$ & $\frac{1}{9}$ & $0$ & $0$ & $-1$ & $0$ & $0$ & $-1$ & $0$ & $0$ & $0$\tabularnewline
\hline 
$\Delta_{2}$ & $0$ & $\frac{1}{9}$ & $0$ & $0$ & $\frac{1}{9}$ & $0$ & $0$ & $\frac{1}{9}$ & $0$ & $0$ & $-1$ & $0$ & $0$ & $-1$ & $0$ & $0$\tabularnewline
\hline 
$\Delta_{1}$ & $0$ & $0$ & $\frac{1}{9}$ & $0$ & $0$ & $\frac{1}{9}$ & $0$ & $0$ & $\frac{1}{9}$ & $0$ & $0$ & $-1$ & $0$ & $0$ & $-1$ & $0$\tabularnewline
\hline 
\end{tabular}
\par\end{centering}
\caption{Effective $U(1)$ symmetries of the SM where we denote $\left\{ U_{1},U_{2},U_{3}\right\} =\left\{ u,c,t\right\} $,
$\left\{ D_{1},D_{2},D_{3}\right\} =\left\{ d,s,b\right\} $ and $\left\{ E_{1},E_{2},E_{3}\right\} =\left\{ e,\mu,\tau\right\} $.\label{tab:SM_eff_symmetries}}
\end{table}

\section{SU(3) anomalous charge with new colored fermions\label{app:SU3_anomalous_charge}}

To have an $U(1)_{X}-SU(3)^{2}$ anomaly, let us consider the option
of introducing new colored fermions. In the KSVZ model \cite{Kim:1979if,Shifman:1979if},
one extends the SM by a pair of vectorlike quarks ${\cal Q}_{L}\sim\left(3,1,y_{{\cal Q}}\right)$,
${\cal Q}_{R}\sim\left(3,1,y_{{\cal Q}}\right)$ and a scalar $\Phi\sim\left(1,1,0\right)$
under the SM gauge group $SU(3)\times SU(2)_{L}\times U(1)_{Y}$ and
the relevant new interaction is
\begin{eqnarray}
-{\cal L} & \supset & \lambda_{{\cal Q}}\Phi\overline{{\cal Q}_{R}}{\cal Q}_{L}+\textrm{H.c.}.\label{eq:KSVZ}
\end{eqnarray}
Since ${\cal Q}_{L,R}$ contribute to the mixed $SU(3)$ anomaly,
eq. (\ref{eq:SU3_anomaly_condition}) becomes
\begin{eqnarray}
U(1)_{X}-SU(3)^{2} & : & 3\left(2q_{Q}^{X}-q_{U}^{X}-q_{D}^{X}\right)+q_{{\cal Q}_{L}}^{X}-q_{{\cal Q}_{R}}^{X}=0,
\end{eqnarray}
which constitute an independent constraint. Imposing four charge conservation
conditions and two mixed $U(1)_{X}-SU(2)_{L}^{2}$ and $U(1)_{X}-SU(3)^{2}$
anomaly-free conditions, we have $9-6=3$ exactly conserved charges
which can be identified with $Y$, $B-L$ and ${\cal Q}$ number where
only ${\cal Q}_{L,R}$ are charged with $q_{{\cal Q}_{L}}^{{\cal Q}}=q_{{\cal Q}_{R}}^{{\cal Q}}$.
Notice that the quark Yukawa interactions, eq. (\ref{eq:KSVZ}) and
the $U(1)_{X}-SU(3)^{2}$ anomaly-free condition force $q_{\Phi}^{X}=0$
and $q_{{\cal Q}_{L}}^{X}=q_{{\cal Q}_{R}}^{X}$. Removing only the
$U(1)_{X}-SU(2)_{L}^{2}$ anomaly-free condition will result in an
anomalous symmetry formed by the linear combination of $B$ and $L$,
which is not equivalent to $B-L$. Removing only the $U(1)_{X}-SU(3)^{3}$
anomaly-free condition, we have a new anomalous symmetry $U(1)_{PQ}$
with $q_{{\cal Q}_{L}}^{PQ}-q_{{\cal Q}_{R}}^{PQ}=q_{\Phi}^{PQ}$
and anomaly coefficient $A_{PQ33}=\frac{1}{2}\left(q_{{\cal Q}_{L}}^{PQ}-q_{{\cal Q}_{R}}^{PQ}\right)=-\frac{1}{2}q_{\Phi}^{PQ}$.
Since the right-handed up quark number $U(1)_{u}$ has an anomaly
coefficient $A_{u33}=\frac{1}{2}q_{u}^{u}$, we can form an anomaly-free
charge combination
\begin{eqnarray}
\overline{PQ} & = & \frac{PQ}{q_{\Phi}^{PQ}}+\frac{u}{q_{u}^{u}}.
\end{eqnarray}
For $T\gtrsim10^{6}$ GeV where interactions mediated by up-quark
Yukawa coupling are out of equilibrium, $\overline{PQ}$ is a conserved
charge. The nonzero charge assignment of fields under $\overline{PQ}$
is shown in Table \ref{tab:PQ_bar_charges}. Without loss of generality,
we can set $q_{u}^{u}=1$.

\begin{table}
\begin{centering}
\begin{tabular}{|c|c|c|c|c|}
\hline 
$ $ & $u$ & ${\cal Q}_{L}$ & ${\cal Q}_{R}$ & $\Phi$\tabularnewline
\hline 
\hline 
$\overline{PQ}$ & $1$ & $\frac{q_{{\cal Q}_{L}}^{PQ}}{q_{{\cal Q}_{L}}^{PQ}-q_{{\cal Q}_{R}}^{PQ}}$ & $\frac{q_{{\cal Q}_{R}}^{PQ}}{q_{{\cal Q}_{L}}^{PQ}-q_{{\cal Q}_{R}}^{PQ}}$ & $1$\tabularnewline
\hline 
\end{tabular}
\par\end{centering}
\caption{$\overline{PQ}$ is a conserved charge for $T\gtrsim10^{6}$ GeV.\label{tab:PQ_bar_charges}}
\end{table}

Depending on the hypercharge $y_{{\cal Q}}$, they can be new interactions
with the SM fields. First of all, these interactions will imply that
the SM fields should carry $PQ$ charges. Second of all, ${\cal Q}$
might cease to be a conserved charge and ${\cal Q}_{L,R}$ can decay
\cite{DiLuzio:2016sbl,DiLuzio:2017pfr}.

\section{SU(2) anomalous charge\label{app:SU2_anomalous_charge}}

To have an $U(1)_{X}-SU(2)_{L}^{2}$ anomaly beyond those of $B$
and $L$, the only option is to introduce new fermions with nontrivial
representation under $SU(2)_{L}$. Let us extends the SM by a pair
of vectorlike $SU(2)_{L}$ doublets ${\cal L}_{L}\sim\left(1,2,y_{{\cal L}}\right)$,
${\cal L}_{R}\sim\left(1,2,y_{{\cal L}}\right)$ and a scalar $\Phi\sim\left(1,1,0\right)$
with the relevant new interaction
\begin{eqnarray}
-{\cal L} & \supset & \lambda_{{\cal L}}\Phi\overline{{\cal L}_{R}}{\cal L}_{L}+\textrm{H.c.}.\label{eq:SU2_model}
\end{eqnarray}
Here as in the SM, the $U(1)_{X}-SU(3)^{2}$ anomaly-free condition
is redundant. Imposing charge conservation conditions and the modified
$U(1)_{X}-SU(2)_{L}^{2}$ anomaly-free condition
\begin{eqnarray}
U(1)_{X}-SU(2)_{L}^{2} & : & 3\left(3q_{Q}^{X}+q_{\ell}^{X}\right)+q_{{\cal L}_{L}}^{X}-q_{{\cal L}_{R}}^{X}=0,
\end{eqnarray}
we have $9-5=4$ exactly conserved charges: $Y$, $B-L$, ${\cal L}$
and $\overline{X}$. Here ${\cal L}$ is the number where only ${\cal L}_{L,R}$
are charged with $q_{{\cal L}_{L}}^{{\cal L}}=q_{{\cal L}_{R}}^{{\cal L}}$
while $\overline{X}$ is the new portal charge. To construct $\overline{X}$,
we note that with $q_{{\cal L}_{L}}^{X}-q_{{\cal L}_{R}}^{X}=-q_{\Phi}^{X}\neq0$,
we have an $U(1)_{X}-SU(2)_{L}^{2}$ anomaly with anomaly coefficient
$A_{X22}=\frac{1}{2}\left(q_{{\cal L}_{L}}^{X}-q_{{\cal L}_{R}}^{X}\right)=-\frac{1}{2}q_{\Phi}^{X}$.
Using $A_{B22}=A_{L22}=\frac{3}{2}$, we can then construct an anomaly-free
charge
\begin{eqnarray}
\overline{X} & \equiv & \frac{X}{q_{\Phi}^{X}}+\frac{1}{3}\frac{c_{B}B+c_{L}L}{c_{B}+c_{L}},
\end{eqnarray}
where $c_{B}$ and $c_{L}$ are any numbers. The nonzero charge assignment
of fields under $\overline{X}$ is shown in Table \ref{tab:X_bar_charges}.
\begin{table}
\begin{centering}
\begin{tabular}{|c|c|c|c|c|c|c|c|c|}
\hline 
$ $ & $Q_{\alpha}$ & $U_{\alpha}$ & $D_{\alpha}$ & $\ell_{\alpha}$ & $E_{\alpha}$ & ${\cal L}_{L}$ & ${\cal L}_{R}$ & $\Phi$\tabularnewline
\hline 
\hline 
$\overline{X}$ & $\frac{1}{9}\frac{c_{B}}{c_{B}+c_{L}}$ & $\frac{1}{9}\frac{c_{B}}{c_{B}+c_{L}}$ & $\frac{1}{9}\frac{c_{B}}{c_{B}+c_{L}}$ & $\frac{1}{3}\frac{c_{L}}{c_{B}+c_{L}}$ & $\frac{1}{3}\frac{c_{L}}{c_{B}+c_{L}}$ & $\frac{q_{{\cal L}_{L}}^{X}}{q_{{\cal L}_{L}}^{X}-q_{{\cal L}_{R}}^{X}}$ & $\frac{q_{{\cal L}_{L}}^{X}}{q_{{\cal L}_{L}}^{X}-q_{{\cal L}_{R}}^{X}}$ & 1\tabularnewline
\hline 
\end{tabular}
\par\end{centering}
\caption{$\overline{X}$ is an exactly conserved charge.\label{tab:X_bar_charges}}
\end{table}

\section{Two Higgs doublets model\label{app:2HDM}}

As we have discussed earlier, another option to have an $U(1)_{X}-SU(3)^{2}$
mixed anomaly is to add another Higgs doublet. Let us consider two
Higgs doublets $H_{u}$ and $H_{d}$ with the following Yukawa interactions
\begin{eqnarray}
-{\cal L} & \supset & y_{u}\overline{U}QH_{u}+y_{d}\overline{D}QH_{d}+y_{e}\overline{E}\ell H_{d}+\textrm{H.c.}.\label{eq:2HD}
\end{eqnarray}
In this case, we have seven fields. Since there are only three conditions
above, we can identify four classically conserved charges $Y$, $B$,
$L$ and $X$ where only $Y$ is anomaly-free while $X$ can have
both $U(1)_{X}-SU(2)_{L}^{2}$ and $U(1)_{X}-SU(3)^{2}$ anomalies.
Imposing the two mixed $U(1)_{X}-SU(2)_{L}^{2}$ and $U(1)_{X}-SU(3)^{2}$
anomaly-free conditions, we are left with $Y$ and $B-L$. With couplings
to the two Higgs doublets as in eq. (\ref{eq:2HD}), the $U(1)_{X}-SU(3)^{2}$
anomaly-free condition is no longer redundant. Furthermore, $\mu^{2}H_{u}H_{d}$
is automatically allowed by the $U(1)_{X}-SU(3)^{2}$ anomaly-free
condition and conversely, $\mu^{2}H_{u}H_{d}$ implies the $U(1)_{X}-SU(3)^{2}$
anomaly-free condition. Indeed, we can write the anomaly coefficient
as
\begin{eqnarray}
A_{X33} & = & \frac{3}{2}\left(2q_{Q}^{X}-q_{U}^{X}-q_{D}^{X}\right)=-\frac{3}{2}\left(q_{H_{u}}^{X}+q_{H_{d}}^{X}\right).
\end{eqnarray}
 The choice $y_{e}\overline{E}\ell H_{u}^{*}$ does not change the
analysis above while having $y_{d}\overline{D}QH_{u}^{*}$ is equivalent
to enforcing the $U(1)_{X}-SU(3)^{2}$ anomaly-free condition.

If we remove only the $U(1)_{X}-SU(2)_{L}^{2}$ anomaly-free condition,
we have an anomalous linear combination of $B$ and $L$ which is
linearly independent to $B-L$. If we remove only the $U(1)_{X}-SU(3)^{2}$
anomaly-free condition, we will have an anomalous $U(1)_{PQ}$. This
is the original PQWW model \cite{Peccei:1977hh,Peccei:1977ur,Weinberg:1977ma,Wilczek:1977pj}
while if we introduce another scalar field $\Phi\sim\left(1,1,0\right)$
with the new interaction $\lambda\Phi^{2}H_{u}H_{d}$, we obtain the
DFSZ model \cite{Zhitnitsky:1980tq,Dine:1981rt}.

The supersymmetrized version has been discussed previously. Just to
recap that out of the three $B-L$ violating terms $U^{c}D^{c}D^{c}$,
$QD^{c}\ell$ and $\ell\ell E^{c}$, we only need to consider $U^{c}D^{c}D^{c}$
and either $QD^{c}\ell$ or $\ell\ell E^{c}$. If we include $\mu H_{u}H_{d}$
term, there is no mixed $U(1)_{X}-SU(3)^{2}$ anomaly, and we have
$7-6=1$ which is the anomaly-free $Y$. Once we remove $\mu H_{u}H_{d}$,
we have $Y$ and $X$ in which the latter has both mixed $U(1)_{X}-SU(2)_{L}^{2}$
and $U(1)_{X}-SU(3)^{2}$ anomalies with anomaly coefficients $A_{X22}=A_{X33}=-\frac{3}{2}\left(q_{E}^{X}-3q_{D}^{X}\right)\neq0$.
The choice $q_{E}^{X}-3q_{D}^{X}=0$ gives the anomaly-free $Y$.
If we further remove $U^{c}D^{c}D^{c}$ ($QD^{c}\ell$ or $\ell\ell E^{c}$)
term, we gain an anomalous $B$ ($L$).

\bibliography{draft}

\end{document}